\documentstyle[aps]{revtex}

\begin{document}
\title{Spinor representation of the general Lorentz Group for spin 1/2
particles and CPT}  
\author{Recai Erdem$^1$}
\addtocounter{footnote}{1}
\footnotetext{E-mail:erdem@photon.iyte.edu.tr}
\maketitle
\small
\begin{center}
Department of Physics\\
Izmir Institute of Technology\\
G.O.P. Bulvari No.16\\
Izmir, Turkey
\end{center}
\normalsize
\begin{abstract}
We show that the attempt to introduce all of the discrete space-time
transformations
into the 
spinor representation of the Lorentz group as wholly independent
transformations (as in the vectorial representation) leads to an
8-component
spinor representation in general. 
The first indications
seem to imply that CPT can be violated in this formulation without going
outside of field
theory. However one needs further study to reach a final conclusion.
\end{abstract}
\section{Introduction}

The question of the implementation of the spinor representation of the
full Lorentz group inculding discrete space-time transformations is
examined in the
excellent study by Wigner [1] and by him and his collaborators [2] and by 
S. Weinberg [3]. Wigner arrived at the conclusion that the
inclusion of time reversal into the spinor representation leads to the
doubling of particle states in general. However he dismissed this
alternative on emprical grounds because the unusual properties of the
mirror particles is not observed among the elementary particles. In spite
of this difficuly there are some studies in the literature on how to
realize these states in the context of field theory. However these studies 
either are specific constructs [4,5] or remain as simple models whose
phenomenological viability are unclear [6]. One needs a phenomenological
scheme which is theoretically well motivated and general enough to embed
different cases as subcases. In this study we shall give such a framework.
Moreover Wigner considers only
usual discrete space-time transformations (essentially CP and T) so that
the doubling of the Hilbert space occur at the level of time reversal
while in this study we inculde charge conjugation in our analysis to
make the formulation more general and to introduce the doubling in the
level of parity. In this way we use the case of parity as a guiding
situtation to derive concerete results for the case of time reversal. By
using this formulation we arrive at the result that the spinor
representation of full Lorentz group leads to 8-component spinor
representation in general. We show that the validity of CPT
theorem in this case is not clear and it needs further study. Moreover in
this way the enlargement of the
Hilbert space is not restricted to the presence of intrinsic parity and
time reversal degrees of freedom in their usual meaning. 
The term,
intrinsic space includes both the intrinsic space in its usual meaning
and the internal space dependence associated with the discrete space-time
transformation so that our
conclusions have a more general applicability. (However we do not consider
projective representation of CPT in order to focuse on time reversal 
so that CPT is equivalent to just
CP$\times$T in this study 
hence
there is a doubling of the Hilbert space rather than a quadrupling [2] in
general. In this respect this study is less general than [1] and [2].) We
also give the outline of
a phenomenological scheme where the mirror particles can not be observed
at the present low energies and this also suppreses a possible CPT
violation which may occur through this formulation. 

The general homogenous Lorentz group is
defined through
its vector representation as the set of transformations, $\Lambda$ [7,8]
\begin{equation}
x^\mu\rightarrow x^{\prime\mu}=(\Lambda x)^\mu=\Lambda^\mu_\rho x^\rho
\label{a1}
\end{equation}
which leaves the quantity $x_\mu x^\mu$ invariant,
\begin{eqnarray}
&&x.x=x_\mu x^\mu=g_{\mu\tau}x^\tau x^\mu=
x_\mu^{\prime} x^{\prime\mu}~,\label{a21} \\
&&\mbox{which implies}~~~
g_{\mu\nu}\Lambda^\mu_\rho\Lambda^\nu_\tau=g_{\rho\tau}
\label{a22}
\end{eqnarray}
where $x=(x^0,x^1,x^2,x^3)$ is an arbitrary vector in 
Minkowski space and $g_{\mu\nu}=diag(1,-1,-1,-1,-1)$ is the 
metric tensor of this space. 
From the Eq.(\ref{a22}) one obtains some conditions on the sign of  
$\Lambda^0_0$ and on the value of det$\Lambda$ which divides the
general homogenous Lorentz
group, $L$ into four sets of transformations  
that are not connected to each other through continous Lorentz
transformations [8].
\begin{eqnarray}
&&L_+^{\uparrow}:~det\;\Lambda=+1,~~~sgn\;\Lambda^0_0=1~~\mbox{which
contains 
I=diag(1,1,1,1)} \nonumber \\
&&L_-^{\uparrow}:~det\;\Lambda=-1,~~~sgn\;\Lambda^0_0=1~~\mbox{which
contains}~~ 
P=diag(1,-1,-1,-1) \nonumber \\
&&L_+^{\downarrow}:~det\;\Lambda=+1,~~~sgn\;\Lambda^0_0=-1~~\mbox{which 
contains}~~ PT=diag(-1,-1,-1,-1) \nonumber \\
&&L_-^{\downarrow}:~det\;\Lambda=-1,~~~sgn\;\Lambda^0_0=-1~~\mbox{which 
contains}~~ T=diag(-1,1,1,1) 
\label{a5}
\end{eqnarray}
where $P$ and $T$ above denote parity and time reversal transformations.

The group of 2-dimensional complex special linear transformations,
$SL(2,C)$ defines another representation of Lorentz group because one can 
represent the 4-dimensional space-time coordinates as a second
rank tensor, $X$ of $SL(2,C)$ under boosts and rotations [9]
\begin{eqnarray}
&&X=\left(\begin{array}{cc}
x_0+x_3&x_1-ix_2\\
x_1+ix_2&x_0-x_3\end{array}\right)~~~~~det\,X=x_\mu x^\mu
\label{a6} \\
&&X^\prime=\left(\begin{array}{cc}
x_0^\prime+x_3^\prime&x_1^\prime-ix_2\prime\\
x_1^\prime+ix_2^\prime&x_0^\prime-x_3^\prime
\end{array}\right)
=X\rightarrow X^\prime=A\;X\;A^\dagger
\label{a7}
\end{eqnarray}
where $A$ is a unimodular $2\times 2$ complex matrix and the primes on
the $x$'s denote their Lorentz transformed form. 
The first rank tensor of $SL(2,C)$, $\chi$ which transforms as
$\chi\rightarrow A\chi$ is used to define fermionic fields.

After obtaining the 
$\Lambda^\mu_\nu$'s in terms of the elements of A by using 
Eq.(\ref{a7}) [10]
one notices that $SL(2,C)$ transformations do not correspond to the 
whole set of the homogenous Lorentz transformations spanned by the group, 
L. 
To be more specific such a analysis reveals that $SL(2,C)$
does not correspond to the whole 
L but to $L_+^\uparrow$.
The correspondance between the elements of $SL(2,C)$ and 
$L_+^\uparrow$ is not a one to one correspondance because to each 
rotation, R belonging to $SO(3)$ subgroup of $L_+^\uparrow$ there corresponds
two unitary matrices $\pm u$ belonging to $SU(2)$ subgroup of $SL(2,C)$. 
Therefore
\begin{equation}
L_+^\uparrow\simeq\frac{SL(2,C)}{Z_2}
\label{a9}
\end{equation}
Moreover there are two
representations of $SL(2,C)$, the dotted and the undotted representions
denoted by $\chi_L$ and $\chi_R$. 

The spinors belonging to the dotted and undotted representations, $\chi_L$
and $\chi_R$, repectively can be parametrized as  
\begin{equation}
\chi_L=\chi~~~~\chi_R\sim i\sigma_2\chi^* \label{a91}
\end{equation}
so that they transform as
[11] 
\begin{eqnarray}
&&\chi_L\rightarrow u_L\chi_L~,~~~
u_L=A=e^{\frac{i}{2}\vec{\sigma}.(\vec{\theta}-i\vec{\nu})}
\nonumber \\
&&\chi_R\rightarrow u_R\chi_R~,~~~u_R=(A^{-1})^\dagger=i\sigma_2
A^*\sigma_2= 
e^{\frac{i}{2}\vec{\sigma}.(\vec{\theta}+i\vec{\nu})}
\label{a13}
\end{eqnarray}
where $\theta_j$ and $\nu_j$ are the rotation and boost parameters along 
$x_j$, respectively.

\section{C, P, T, and spinor representation of Lorentz group}
\subsection{The Standard Formulation}

In the absence of any (explicit) internal
space dependence (so that $\chi_L$, $\chi_R$ can be wholly expresed 
only in terms of their space-time dependences) one can, for example,
parametrize an abitrary pair of $\chi_L$ and $\chi_R$ as 
\begin{eqnarray}
&&\chi_L=e^{\frac{i}{2}\vec{\sigma}.(\vec{\theta}_{01}-i\vec{\nu}_{01})}
\left(\begin{array}{c}
0\\
1\end{array}\right) \nonumber \\
&&\chi_R=e^{\frac{i}{2}\vec{\sigma}.(\vec{\theta}_{02}+i\vec{\nu}_{02})}
\left(\begin{array}{c}
0\\
1\end{array}\right) \label{a15}
\end{eqnarray}

In the massless case all the invariants of the form $x_\mu x^\mu$ are
identically equal to zero
\begin{equation}
x_\mu x^\mu=det X=0 \label{a151}
\end{equation}
One can take two specific $X$'s; 
$X_1=\chi_L\chi_L^\dagger$, $X_2=\chi_R\chi_R^\dagger$. Of course, $det
X_1$=$det X_2$=0, hence
\begin{equation}
det(X_1X_2)=0=-\chi_R^\dagger\chi_L\chi_L^\dagger\chi_R \Rightarrow
\chi_R^\dagger\chi_L=\chi_L^\dagger\chi_R=0
\label{a152}
\end{equation}
This sets the form of the spinors (up to a constant) as
\begin{eqnarray}
&&\chi_L=e^{\frac{i}{2}\vec{\sigma}.(\vec{\theta}_{0}-i\vec{\nu}_{0})}
\left(\begin{array}{c}
0\\
1\end{array}\right) \nonumber \\
&&\chi_R=e^{\frac{i}{2}\vec{\sigma}.(\vec{\theta}_{0}+i\vec{\nu}_{0})}
\left(\begin{array}{c}
1\\
0\end{array}\right) \label{a154}
\end{eqnarray}
Therefore the left-handed and right-handed spinors in the massless case
are not related to each other just as their space reflected form. They are
simply related by
\begin{equation}
\chi_R =i\sigma_2\chi^* \label{a155}
\end{equation}
One observes that $\chi_L$ and $\chi_R$ are related by time reversal 
and the simple space reflection does not define a physical spinor.
There is only one independent discrete space-time transformation,
$C^\prime$ (for the spinor part of the fermion field) which can be
interpreted as just the usual time reversal 
\begin{equation}
C^\prime=i\sigma_2 K~,~~~ S\simeq C^\prime = T
\label{a23}
\end{equation}
where $K$ denotes complex conjugation; $S$ and $T$ denote the usual
space-reflection and time reversal,
respectively.
Therefore, in the massless case, the Lorentz
group itself specifies the right-handed 2-component spinor,
$\chi_R$ 
corresponding to the time reversal of a particular left-handed one,
$\chi_L$ out of the family
of dotted and undotted spinors in Eq.(\ref{a15}). We should remark that
although $C^\prime$ acts on the same Hilbert space as $SL(2,C)$
transformations it is not a member of $SL(2,C)$ thus it extends $SL(2,C)$. 
Although the group which includes the discrete
space-time transformations (and acts on the spinors) is given by
\begin{equation}
\frac{SL(2,C)}{Z_2}\otimes D^\prime~~,~~~~D^\prime=(1,C^\prime)
\label{a231}
\end{equation}
It is effectively equivalent to $SL(2,C)$ if the dotted and undotted
representations are not coupled in the Lagrangian because in that case
either the Lagrangian is automatically invariant under $C^\prime$ if there
is no internal space differing the dotted and undotted representations or
$C^\prime$ is equivalent to an internal space transformation if the dotted
and undotted representations differ by an additional internal space
dependence.

We have seen above that one can use $C^\prime$ in Eq.(\ref{a23}) to
specify the particular right-handed partner, $\chi_R$ of a particular
left-handed 2-component spinor, $\chi_L$. Moreover in this way one
guarantees that $\chi_L$ and $\chi_R$ transform as left-handed and right
spinors as in Eq.(\ref{a13}). Another way to specify a particular
$\chi_R$ corresponding to to a particular $\chi_L$ 
(out of the family of right handed spinors
transforming as in
Eq.(\ref{a13}) 
so that their
correct transformation
properties as in Eq.(\ref{a13}) are guaranteed) is to introduce a fermion
mass term for the Lagrangian  
\begin{equation}
{\cal L}_m=m(\chi_L^\dagger\chi_R+\chi_R^\dagger\chi_L)
\label{a2311}
\end{equation}
In fact the general transformation rules of $\chi_L$, $\chi_R$ given in 
Eq.(\ref{a13}) embodies the massive Dirac equation in general (and the
massless case is a special case where (\ref{a2311})
identically vanishes due to the particular relation between $\chi_L$ and
$\chi_R$ as discussed above). One can start from the value of $\chi_L$,
$\chi_R$ at a particular reference frame and the transformation properties
of $\chi_L$, $\chi_R$ under $SL(2,C)$ to derive the Dirac equation 
[5, 12 - 15].
The
fact that $\chi_L$ and $\chi_R$ are coupled (so that they form a
(4-component spinor) system moving together) simplifies the form of
$\chi_L$ and $\chi_R$ ($\nu_{01}=\nu_{02}$)
\begin{eqnarray}
&&\chi_L=e^{\frac{i}{2}\vec{\sigma}.(\vec{\theta}_{01}-i\vec{\nu}_{0})}
\left(\begin{array}{c}
0\\
1\end{array}\right) \nonumber \\
&&\chi_R=e^{\frac{i}{2}\vec{\sigma}.(\vec{\theta}_{02}+i\vec{\nu}_{0})}
\left(\begin{array}{c}
0\\
1\end{array}\right) \label{a1511}
\end{eqnarray}
By making the use of Eq.(\ref{a13}) one gets 
\begin{eqnarray}
&&\chi_L=e^{\vec{\sigma}.\vec{\nu}}\chi_R=
(\cosh\nu+\vec{\sigma}.\hat{\nu}\sinh\nu)\chi_R=
\frac{1}{m}(p_0+\vec{\sigma}.\vec{p})\chi_R
\nonumber \\
&&\chi_R=e^{-\vec{\sigma}.\vec{\nu}}\chi_L=
(\cosh\nu-\vec{\sigma}.\hat{\nu}\sinh\nu)\chi_R=
\frac{1}{m}(p_0-\vec{\sigma}.\vec{p})\chi_L
\label{a19}
\end{eqnarray}
which becomes the Dirac equation after the substitution 
$p_\mu\rightarrow i\partial_\mu$.  

Eq.(\ref{a1511}) implies that there can be more than one $\chi_R$
corresponding to a particular $\chi_L$ in the absence of an intrinsic
space dependence for the fermions in addition to the usual space-time
dependence.
Moreover one can differ the left-handed and right-handed fermions through
internal space representations as in the standard model. This means that
in general the fermions carry an intrinsic space
dependence. 

If one does not want to have an intrinsic space dependence the
only way to make the correct match between $\chi_L$ and $\chi_R$
with correct transformation property under $SL(2,C)$ is to take
$\chi_R=i\sigma_2\chi_L^*$. The resulting 4-component spinor is
the Majorana spinor
\begin{eqnarray}
\left(\begin{array}{c}
\chi\\
i\sigma_2\chi^*\end{array}\right)
\label{a1521}
\end{eqnarray}
where both components of $\chi$ are non zero in contrary to the situation
in Eq.(\ref{a154}). There is only one independent discerete transformation
\begin{equation} 
P=\gamma^0~,~~
S\simeq P \simeq e^{i\beta}T
\label{a24}
\end{equation}
where $\beta$ is an unobservable phase factor and $S$ stands for the 
usual space reflection (which is a specific case of $P$). 
In this particular case one can write P in an equivalent way to
Eq.(\ref{a24})
\begin{eqnarray}
P=C^{\prime\prime}~=\left(\begin{array}{cc}
i\sigma_2K&0\\
0&-i\sigma_2K\end{array}\right)
\label{a241}
\end{eqnarray}
Although in this case the spinor 
apparentely is 4-component, in fact, it is equivalent to a 
single 2-component theory since the Dirac equation reduces to a 
single equation in terms of a 2-component spinor and its complex 
conjugate. Therefore the situation is the same as
the massless case with respect to the number of independent discrete 
transformations accomodated in the spinor representation of the Lorentz 
group. The relevant group for the spinor representation of the 
Lorentz group is given by
\begin{equation}
\frac{SL(2,C)}{Z_2}\otimes D~~,~~~D=(1,P)
\label{a25}
\end{equation}
because $\chi_L$ and $\chi_R$ are related by $P$ through the mass term.
Here $D$ denotes a $Z_2$ group which is generated by
$C^{\prime\prime}$ in Eq.(\ref{a241}) and so effectively by the
elements
$(1,~i\sigma_2 K)$ where $K$ denotes the complex conjugation acting only
on the spinor part of fermion field. However since $i\sigma_2K$ simply
generates the transformation from the fundamental representation to the
conjugate representation the group generating the spinor representation is
effectively $SL(2,C)$ in this case as well.

We had seen in the previous paragraphs that in general the relation
between $\chi_L$ and $\chi_R$ is
not as simple as the massless case and the Majorana case. There can be an
additional source of difference between $\chi_L$ and $\chi_R$ in addition
to their tansformation rules under $SL(2,C)$, namely the difference
between their form at their rest frame as can be infered from 
Eq.(\ref{a1511}).
Moreover $\chi_L$ and $\chi_R$ may differ in their internal group content
as in the case of the standard model. So one should include these type of
intrinsic space dependences through an additional $Z_2$ transformation
acting on the intrinsic space in addition to the usual $SL(2,C)$
transformations when one wants to relate $\chi_L$ and $\chi_R$. 
The general 4-component spinor can be parametrized as [15]
\begin{eqnarray} 
\psi=\left(\begin{array}{c} \psi_1(p_2)\\ \psi_2(p_1)\end{array}\right)~,
\psi=\left(\begin{array}{c}
\chi_L\\
\chi_R\end{array}\right)
\label{a26}
\end{eqnarray}
However a single 4-component spinor notation is not the accurate 
tool in this case   
because the left-handed and right-handed components couple differently in 
general (for example through gauge interactions). So the correct procedure
is to define two 4-component spinors 
defined by
\begin{eqnarray} 
\Psi_L=\left(\begin{array}{c} 
\psi_1(p_2)\\ 0\end{array}\right)
=\left(\begin{array}{c}
\chi_L\\
0\end{array}\right)~,~~
\Psi_R=\left(\begin{array}{c} 
0\\\psi_2(p_1)\end{array}\right)
=\left(\begin{array}{c}
0\\
\chi_R\end{array}\right)
\label{a27}
\end{eqnarray}
and the operation of 
parity is defined by
\begin{eqnarray} 
&&P: \Psi_L=\left(\begin{array}{c}
0\\
\psi_2(p_1)
\end{array}\right)
=\left(\begin{array}{c}
0\\
\chi_R\end{array}\right)=\Psi_R \nonumber \\
&&P: \Psi_R=\left(\begin{array}{c}
\psi_1(p_2)\\
0
\end{array}\right)
=\left(\begin{array}{c}
\chi_L\\
0\end{array}\right)=\Psi_L \nonumber \\
\label{a277}
\end{eqnarray}
where $p_1$, $p_2$ stand for the space-time part of $\psi$ and denote the 
dotted and undotted representations of $SL(2,C)$, the subscripts $1$ and 
$2$ denote the intrinsic space dependence. 

The generalization of the Majorana spinor through the introduction of the 
intrinsic degree of freedom results in a new nontrivial discrete 
transformation (in addition to $P$), charge conjugation, $C$ given by,
\begin{equation}
C:\Psi_{L(R)}=\gamma^2\Psi_{L(R)}^*
\label{a271}
\end{equation}
The symetries related to both $P$ and $C$ are observable because one can 
couple the same gauge field $\not{B}$ both to 
$P:\Psi_{L(R)}$ and $\Psi_{L(R)}$
($C:\Psi_{L(R)}$ and $\Psi_{L(R)}$) in a Lorentz invariant way to detect
any possible violation of the symmetry. 

When the intrinsic space dependence is omitted (i.e. in the Majorana 
case) $C$ becomes equivalent to identity because in that case $\Psi_L$ 
and $\Psi_R$ can be combined in a single 4-component spinor 
$\psi$ of the form of Eq.(\ref{a1521}) since there is no intrinsic space
differentiating 
$\chi_L$, $\chi_R$ (for example through gauge intercations) and hence
\begin{equation}
C:\psi=\psi \label{a28}
\end{equation}
in the case of Majorana spinors.This reveals that the emergence of $C$ as
a nontrivial additional 
independent discrete transformation is related to the introduction of the 
intrinsic parity degree of freedom. $T$ is found to be equivalent to $CP$ 
up to an unobservable phase factor and a gamma matrix in the case of 
spinors and
they are wholly equivalent for the physical obsevables. In this case the
relevant group for the spinor representation of the Lorentz group is
\begin{equation}
\frac{SL(2,C)}{Z_2}\otimes \frac{S^\prime\otimes Z_2^\prime}{Z_2}
~~,~~~S^\prime=(1,S)
\label{a281}
\end{equation}
where $S$ stands for the ordinary space reflection and $Z_2^\prime$ 
stands for the intrinsic space. The $Z_2$ term in the denominator of 
$S\otimes Z_2^\prime$ is introduced in order to avoid fermion doubling 
[16,17] so that the dotted (undoted) representation of $SL(2,C)$ is
associated with the subscript $2(1)$ in Eq.(\ref{a27}). One should be
careful
in handling 
Eq.(\ref{a281}) because $Z_2^\prime$ does not act on the whole fermionic 
field but only on its space-time part. The fact that $CP$ is a good 
symmetry of nature combined with the above group structure explains ( at 
least at presently attainable relatively low energies) the rather restricted 
form of the physical Lagrangian, for example, the possibility of 
obtaining the general form of the standard model Lagrangian under the 
requirement of an intrinsic parity invariance [18].
The conclusion of the above analysis be summarized as
follows: 
{\bf If one wants to accomodate more than one discrete space-time 
transformation into the spinor representation of Lorentz group one must 
introduce (an) intrinsic degree(s) of freedom accomponying the corresponding
discerete discrete space-time transformations and one should incerease
the number of the components of the spinor accordingly}. 
In fact mathematically this not
an 
unexpected result. Under the group $SL(2,C)$ which corresponds to spinor 
representation of Lorentz group all discrete space-time transformations 
are equivalent, that is, they induce interchange of the dotted and 
undotted representations of $SL(2,C)$. In order to differentiate them as 
in the vectorial representation of the general Lorentz group one must 
introduce extra (intrinsic) degrees of freedom.

\subsection{Full Implementation of the Discrete Space-Time Transformations
in the Spinor Representation: 8-Component Formalism}

One can follow the same procedure in the case of parity for 
time reversal to introduce it as an independent discrete space-time 
transformation in the context of the spinor representation of the Lorentz 
group. We extend the definition of time reversal, up to an observable 
phase, as the usual time reversal
followed by an intrinsic time reversal transformation ( 
the intrinsic parity degree of freedom can identified, for example, with the 
different transformation properties of $\psi_1$ and $\psi_2$ under the 
internal group as in the standard model in the case of parity), that is, 
\begin{equation}
{\cal T}:\psi_{1(2)}(p_{1(2)})\rightarrow T:\psi_{2(1)}(p_{1(2)})
\label{a29}
\end{equation}
up to some phase factor and we couple $\psi_1$, $\psi_2$  through the
following interaction
\begin{eqnarray}
&&{\cal L}= m\bar{\psi}_1\psi_1+m\bar{\psi}_2\psi_2+M\psi_1^\dagger\psi_2+
M\psi_2^\dagger\psi_1=\Psi^{\dagger} M^\prime\Psi \label{a30} \\
&&M^\prime=\left(\begin{array}{cc}
\gamma^0m&M\\
M&\gamma^0m\end{array}\right)~~~\Psi=\left(\begin{array}{c}
\psi_1\\
\psi_2\end{array}\right)
\end{eqnarray}
After the inclusion of the fermion kinetic term the Lagrangian becomes
\begin{eqnarray}
&&{\cal L}= 
\bar{\Psi}D_\mu\Gamma^\mu\Psi
+\bar{\Psi} \tilde{M}\Psi \label{a301} \\
&&\tilde{M}=\left(\begin{array}{cc}
m&\gamma^0M\\
\gamma^0M&m\end{array}\right)~~~~\Gamma^\mu=
=\left(\begin{array}{cc}
\gamma^\mu&0\\
0&\gamma^\mu\end{array}\right)
\end{eqnarray}
which is written in the 8-component spinor formulation [16] (the 
convention for $\Gamma^\mu$ matrices here is different than the one in
Ref.16
but they can be turned to each other). 
The diagonalization of $\tilde{M}$ results in two 4-component spinors 
$\psi$,
$\psi^{c}$,
\begin{equation}
\psi\propto M\psi_1-m\gamma^0\psi_2~,~~
\psi^c\propto m\gamma^0\psi_1+M\psi_2
\label{a31}
\end{equation}
So the physical fermions are the combination of a set of fermions with 
the charge conjugates of another set of fermions. If one identifies
$\psi_1$ with usual particles then the conservation of electric charge
allows the mixing in Eq.(\ref{a31}) only for neutrinos. The
extended time reversal, ${\cal T}$ can be written explicitly as 
\begin{eqnarray}
{\cal T}=\left(\begin{array}{cc}
0&e^{i\eta}\,T\\
e^{-i\eta}\,T&0\end{array}\right)~~\mbox{where}~~T=i\gamma^1\gamma^3
\label{a311}
\end{eqnarray}
Here $T$ denotes the usual time reversal transformation. 
The above form of ${\cal T}$ which is introduced by S. Weinberg [3]
includes the two cases, ${\cal T}^2=1$ and ${\cal T}^2=-1$ in [1,2] as
subcases. 
${\cal T}^2=1$ and  
${\cal T}^2=-1$ correspond to $e^{i\eta}=1$ and $e^{i\eta}=i$,
respectively. Although the operator ${\cal T}^2$ generate superselection
rules for ${\cal T}^2=\pm 1$ (becasue it commutes with all operators) This
is not true for arbitrary value of $\eta$ because in that case ${\cal
T}^2$ is not proportional with unit matrix in general. This allows mixing
of the
particle states belonging to the Hilbert spaces denoted by the subscripts
$1$ and $2$ in Eq.(\ref{a30}). This mixing, in
turn, makes it impossible to absorb the phase $\eta$ into redefinition
$\psi_1$ and $\psi_2$. 
The relevant group 
for the spinor representation in this case is generated by the following
class of transformations
\begin{equation}
\frac{SL(2,C)}{Z_2}\otimes 
\frac{S\otimes Z_2^\prime}{Z_2}
\otimes\frac{T\otimes Z_2^{\prime\prime}}{Z_2}
\label{a282}
\end{equation}
where $T$ stands for the usual time reversal and $Z_2^{\prime\prime}$ 
stands for the intrinsic time reversal degree of freedom. The
corresponding groups for the subcase ${\cal T}^2=\pm 1$ (and ${\cal
CPT}^2=\pm 1$) are derived in Ref.1.
The explicit form of the relevant group in the present case depends on the
value
of ${\eta}$ and this point remains outside the main objective of this
study.

We want to make a comment at this point: In the light of the present 
study the emergence of a 8-component 
formulation in the construct of D.V. Ahluwalia for neutral spin 1/2 
particles [5,16,19] through the careful analysis of Ref.16 is not 
surprising. As we have seen the only
possible 4-component spinor which can be constructed without the use of an 
intrinsic space is the Majorana spinor. The modification of Majorana 
spinor even with introduction of a phase factor implies differing of  
$\chi_L$ and $\chi_R$ by an intrinsic space dependence in addition to the 
usual space reflection or time reversal. In the Ahluwalia's construct
the upper and lower components of the (true) original Majorana spinor are
related by the usual time reversal
and then a intrinsic space dependence differentiating self-conjugate and
anti-self-conjugate (Majorana) spinors is introduced and finally an
intrinsic parity
degree of freedom is introduced to double the number of 4-component
spinors so that the most convenient tool to handle this formulation is
8-component spinor representation. This construct can be understood as a
special case of the general formulation given here because there are only
two differences between these constructs: i) the order 
of introducing intrinsic parity and intrinsic time reversal are different.
ii) the form of intrinsic time reversal in Ahluwalia's construct has a
rather specific form while the intrinsic time reversal introduced here has
a general form.

\section{8-component formulation and CPT}

The Lagrangian in Eq.(\ref{a30}) is invariant under the interchange 
of the subscripts 1 and 2. One can also take the other terms of the full 
Lagrangian invariant under this interchange. In this case T transformation 
will be equivalent to time reversal 
so that CPT theorem necessarily holds. However one can make the 
Lagrangian non-invariant under this interchange, for example, by taking 
the mass term in front of 
$\bar{\psi}_1\psi_1$ 
to be different than the 
one in front of $\bar{\psi}_2\psi_2$
\begin{equation}
{\cal L}=
m\bar{\psi}_1\psi_1+m^\prime\bar{\psi}_2\psi_2+M\psi_1^\dagger\psi_2+
M\psi_2^\dagger\psi_1=\Psi^{\dagger} M^\prime\Psi \label{a32}
\end{equation}
In this case CPT invariance is 
not automatic as we shall see below. 

CPT theorem states that the condition of weak local commutivity for
fields (which is satisfied by all reasonable fields) is enough for the
following equation to be satisfied [8]
\begin{equation}
(\Psi_0, \phi_\mu(x_1)\phi_\nu(x_2).......\phi_\rho(x_n)\Psi_0)=
i^F(-1)^J(\Psi_0,
\phi_\mu(-x_1)\phi_\nu(-x_2).......\phi_\rho(-x_n)\Psi_0)
\label{a33}
\end{equation}
(where $F$ stands for the number of Majorana type fermion bilinears $J$ is
the
magnitude of angular momentum;
$\phi_\eta(x_k)$ is the field operator
(or a fermion bilinear in the case of fermion) at the point $x_k$; the
subscripts $\mu,\nu,..$ stand for possible additional degrees of freedom
for the fields ( or bilinears) as the components of vectors for vector
fields.) provided the fields satisfy the following axioms  \\
I. For each test function $f$ defined on space-time, there exists a set 
of $\phi_1(f),\phi_2(f),...,\phi_n(f)$ and their adjoints which are 
defined on a
domain $D$ of vectors dense in the Hilbert space, ${\cal H}$. Furthermore
$D$ is a linear space containing $\Psi_0$.
All
the vectors obtained after the application of (i- the Lorentz group
transformations, ii- the field operators) are in the physical Hilbert
space, ${\cal H}$. Moreover for all $\Phi,\Psi\in D\subset {\cal H}$
and $\phi_\mu(f)$ being a field defined as a functional of the test
function $f$, $(\Phi,\phi_\mu(f)\Psi)$ is a tempered distribution. \\
II. The equation
\begin{equation}
U(\Lambda,a)\phi_j(f)U(\Lambda,a)^{-1}=\sum
S_{jk}(a^{-1})\phi_k(\{\Lambda,a\}f)
\label{a34}
\end{equation}
is satisfied; here $\{\Lambda,a\}f=f(a^{-1}(x-a)$ and $U(\Lambda,a)$ is
the unitary representation of Poincare Group with $\Lambda$ corresponding
to the homogenous Lorentz group and $a$ to the translations. \\

In the case of 4-component spinors CPT invariance reduces to
the validity of Eq.(\ref{a33}). This can be seen as follows: Under the
discrete space-time transformations fermions transform as [20]
\begin{eqnarray}
P:\psi(\vec{x},t)\rightarrow e^{i\beta_P}\gamma^0\psi(-\vec{x},t)~`~~
C:\psi(\vec{x},t)\rightarrow ie^{i\beta_C}\gamma^2\psi^*(\vec{x},t)~`~~
T:\psi(\vec{x},t)\rightarrow ie^{i\beta_T}\gamma^1\gamma^3\psi(-\vec{x},t)
\nonumber \\
CPT:\psi(\vec{x},t)\rightarrow
-i\gamma_5\psi^*(-\vec{x},-t)=
-i\,e^{i(\beta_C-\beta_P-\beta_T)}
\gamma^0\gamma_5\bar{\psi}^T(-\vec{x},-t)
\label{a35}
\end{eqnarray}
here $\psi$ denotes the whole fermionic field ( not just the spinor
part on the contrary to the previous sections). The fermion bilinear
$X^{AB}=\bar{\psi}^A\Gamma\psi^A$, under $CPT$, transforms as
\begin{eqnarray}
&&CPT: \bar{\psi}^A\Gamma\psi^B\rightarrow \bar{\psi}^B\Gamma^\prime\psi^A
\label{a36} \\ 
&&
\Gamma^\prime=\Gamma~~~\mbox{for}~~\Gamma=1,i\gamma_5,\sigma_{\mu\nu}=
[\gamma_\mu,\Gamma_\nu] \nonumber \\
&&
\Gamma^\prime=-\Gamma~~~\mbox{for}~~\gamma_\mu,\gamma_5\gamma_\mu
\nonumber
\end{eqnarray}
where the superscripts $A$, $B$ denote different species of fermions.
After contraction of $X^{BA}$ with itself or with other $X^{AB}$'s
and comparing it with Hermitian conjugate of the original terms
result in Eq.(\ref{a33}). One can perform a similiar and simpler
procedure for the other type of fields (i.e. for scalars, vector fields,
etc.) to get the same conclusion. So in the usual 4-component spinor case
the validity of Eq.(\ref{a33}) is equivalent to CPT invariance. However in
the case of extended-T
\begin{equation}
CPT:\Psi(\vec{x},t)\rightarrow
-iL^0e^{i(\alpha_P+\alpha_C+\alpha_T)}\gamma^0\gamma_5\bar{\Psi}^T
(-\vec{x},-t)~,~~~L^0=\left(\begin{array}{cc}
0&e^{i\eta}I_4\\
e^{-i\eta}I_4&0\end{array}\right)
\label{a37}
\end{equation}
One can see from the above derivation that CPT invariance is not
equivalent to the validity of Eq.(\ref{a33}) in this case. For example
\begin{equation}
CPT:\bar{\Psi}^A\tilde{M}\Psi^B\rightarrow 
\bar{\Psi}^BL^0\tilde{M}L^0\Psi^A
\label{a38}
\end{equation}
So the requirement of the equality of the Hermitian conjugate of
Eq.(\ref{a38}) to the original term does not imply Eq.(\ref{a33}). If one
takes $T$ as the usual time reversal one obtains Eq.(\ref{a33}) but then
the hypothesis II., Eq.(\ref{a34}) is not satisfied because $\psi_1$ mixes
with $\psi_2$ so $\psi_1$ does not span the whole physical Hilbert space
so that
one can not guarantte a Fourier series in terms of $\psi_1$ is enough to
experess the left hand side of Eq.(\ref{a34}). Therefore it seems that CPT
theorem does not hold in this case. 

One can understand the above conclusion through a more physical argument.
After one introduces an intrinsic parity a $PT$ transformation (on the
spinor part of the fermion field) in the
spinor representation of the Lorentz group is not equivalent to identity
because $P$ is not the simple space reflection anymore. However one can
introduce the charge conjugation, $C$ so that $CP$ is effectiveley
equivalent
to space reflection so that $CPT$ on the spinor part of the fermion field
is equivalent to identity. Therefore, in anology with the 4-component
case, 
after introduction of an intrinsic
time reversal degree of freedom one expects to have a $\tilde{C}$
transformation so that $CPT\tilde{C}$ is the exact symmetry of nature.
However in order to arrive to a final conclusion on CPT invariance of
8-component spinor representation introduced here one needs further, more
rigorous and more detailed annalysis on the subject.  
A more detailed study to check CPT 
theorem in this case will be an important task to have a better insight 
on the subject in future. Another point to mention is that (without an
ad hoc CPT invaraince requirement) the simplest 
scheme
consistent with the experimental bounds on CPT violation is to take
$m^\prime$ in Eq.(\ref{a32}) to be very heavy
and $M$ sufficiently small so that at low
energies all the forces interact with the mirror particles, $\psi_2$
extermly
weekly. In this way one can not observe the additional degree of freedom
associated with the extended time reversal so that CP${\cal T}$
transformation is
equivalent to the usual CPT transformation and thus CPT 
is essentially
conserved. All these points neeed further study to have a clear picture.

\section{Conclusion}

In this study we have shown that only one independent discrete space-time 
transformation (for example parity) can be formulated in addition to the 
continous Lorentz 
transformations in the spinor representation of the Lorentz group. The 
attempt to accomodate the other discrete space time transformation (for 
example time reversal) as an independent transformation in the spinor 
representation requires the introduction of furhter intrinsic degrees of 
freedom and this leads to an 8-component spinor notation as the most 
convenient choice. The first indications 
seem to imply that it may be possible to violate CPT in this scheme 
without going outside of field theory. However a final decision requires 
a further rigourous study on the subject. In any case such a study will 
be useful to get a better insight to the problem of accomodation of 
discrete space-time transformations in the spinor representation of the 
general Lorentz group. We also believe that the studies in this direction
will also be useful to have a deeper understanding of the notion of
mirror particles [21]. Another point to be addressed in the future studies
is the 
geometrization of the time reversal as done in Ref.15 in order to get 
the all aspects of the extended time reversal defined in this study  
in a complete and consistent way.

\acknowledgements
I would like to thank Professor R.N. Mohapatra for his
encouragements to present my findings on the topic.


\begin{thebibliography}{99}
\bibitem{} E.P. Wigner, in {\it Group Theoretical Concepts and Methods in
Elementary Particle Physics, Lectures of the Istanbul Summer School of
Theoretical Physics}, edited by F. G{\"u}rsey (Gordon and Breach Science
Publishers, New York, 1964)
\bibitem{} R.M.F. Houtappel, H. Van Dam and E.P. Wigner, Rev.Mod.Phys.
{\bf 37}, 595 (1965)
\bibitem{} S. Weinberg, \it{The Quantum Thoery of Fields} Vol.I
Foundations \rm (Cambridge University Press, 1995), Appendix C of Section
2.
\bibitem{} D.V. Ahluwalia, M.B. Johnson and T. Goldman, Phys.Lett. B {\bf
316}, 102 (1993)
\bibitem{} D.V. Ahluwalia, Int.J.Mod.Phys. {\bf{A 11}}, \rm
1855 (1996)
\bibitem{} J. Giesen, preprint, hep-th/9410236
\bibitem{} W. R{\"u}hl, \it{Lorentz Group and Harmonic Analysis} \rm (W.A. 
Benjamin, New York, 1970)
\bibitem{} R.F. Streater, A.S. Wightman \it{PCT, Spin and Statistics, and 
All That} \rm (W.A. Benjamin, New York, 1964)
\bibitem{} M.A. Naimark, \it{Linear Representations of the Lorentz
Group} \rm
(Pergamon, London, 1964)
\bibitem{} For the details see Chapter 1 of Ref.7
\bibitem{} P. Ramond, \it{Field Theory: A Modern Primer} \rm
(Benjamin/Cummins, Reading Mass., 1990)
\bibitem{} L.H. Ryder, \it{Quantum Field Theory} \rm (Cambridge Univ.
Press, 1986)
\bibitem{} F.H. Gailo and T.G. Alvarez, {\it{Am.J.Phys}} {\bf{63}}, \rm 
177 (1995); 
hep-th/9807211
\bibitem{} V.V. Dvoeglazov, {\it{Hadronic J. Suppl.}} {\bf{10}}, \rm
349 (1995)
349
\bibitem{} R. Erdem, {\it{Mod.Phys.Lett}} {\bf{A 13}}, \rm 465 (1998) 
\bibitem{} V.V. Dvoeglazov, {\it{Nuovo Cim.}} {\bf{A 108}}, \rm 1467 
(1995)
\bibitem{} G. Ziino, {\it{Int.J.Mod.Phys.}} {\bf{A 11}}, \rm 2081 (1996)
\bibitem{} R. Erdem, {\it{Phys.Rev.}} {\bf{D 52}}, \rm 3072 (1995) 
\bibitem{} V.V. Dvoeglazov, {\it{Mod.Phys.Lett}} {\bf{A 12}}, \rm
2741 (1997)
\bibitem{} J.D. Bjorken and S.D. Drell, \it{Relativistic Quantum Fields}
\rm (McGraw-Hill, USA, 1965)
\bibitem{} Z.G. Berezhiani and R.N. Mohapatra, {\it{Phys.Rev.}} {\bf{D
52}}, \rm
6607 (1995);\\
R. Foot and R. Volkas, \it{Phys.Rev.} \bf{D 52}, \rm 6595 (1995);\\
B. Brahmachari and R.N. Mohapatra, hep-ph/9805429

\end{thebibliography}
\end{document}